# Deep Machine Learning in MIMO Communication Systems

Mohammad Reza Ghavidel Aghdam, and Alireza Naghavi Faculty of Electrical and Computer Engineering, University of Tabriz, Tabriz, Iran.
Emails: Ghavidel1992@tabrizu.ac.ir, Alireza_nagavi@yahoo.com

**Abstract**

This paper presents an innovative approach to enhancing machine learning-based communication systems, specifically focusing on multiple-input multiple-output (MIMO) configurations using autoencoders. We optimize the transmitter, receiver, and channel simultaneously under conditions of noise and channel fading, aiming to minimize the bit error rate (BER). By incorporating the Rayleigh fading channel—a widely recognized model for wireless channel impairments—into the autoencoder framework, we directly train the communication system to handle real-world conditions. We introduce a novel optimization process tailored for deep learning-based MIMO communication, and thoroughly analyze the resulting BER performance across various signal-to-noise-ratio (SNR) levels. Our simulation results reveal that the proposed end-to-end wireless communication system achieves significantly lower BER compared to conventional block-based processing methods, highlighting its potential for more efficient and reliable wireless communication.

## I. Introduction

The fields of communication technology and machine learning are converging [1]–[4]. Today's communication systems create a massive amount of traffic data, that can assist to enhance the management and design of networks and communication components significantly when they are combined with advanced machine learning methods [1], [5], [6]. Machine learning methods are of great importance in emerging application fields of communication technology such as next-generation wireless communication or the Internet of things (IoT) [7].

The current limitations of Long Term Evolution (LTE) and LTE-Advanced (LTE-A), like latency, reliability, capacity, and some of the necessities should be considered in next-generation wireless communication networks. The state-of-the-art literature for next-generation wireless networks such as: providing higher network energy efficiency, growth in traffic, providing better Quality of Service (QoS), supporting a wide range of applications and providing improved security and privacy [8], [9]. deploying intelligence in the network is one of the possible approaches to solve these limitations [10].

In recent years, Machine Learning (ML) utilized in communication systems [11]. ML algorithms can be simply classified as supervised, unsupervised and reinforcement learning (RL). The adjective supervised specifies if there are training samples in the database or not [12] and RL appeared as a new classification that was inspired by behavioral psychology [13]. Recently, an end-to-end learning-based approach to physical layer communication system design was introduced in [14]. In this system, a channel autoencoder-based deep learning (DL) technique for the transmitter, receiver, and signal representation optimization is used.

Several new uses of DL in the physical layer of communication systems are discussed in [15]. By interpreting a communication system as an autoencoder, they investigated a fundamentally new approach to think about communication system designs as an end-to-end reconstruction task that attempts to jointly enhance transmitter and receiver elements in a single process. In [16], a fundamental idea of the DL and its use to wireless communication systems is demonstrated through examining the resource allocation scheme based on a deep neural network (DNN) in which various goals with various constraints can be satisfied through the end-to-end deep learning.

Considering that the RA can be formulated into a non-convex problem, the iterative methods based on Lagrangian relaxation are proposed in [17]. Nevertheless, this approach needs several iterative computations, which possibly lengthens the computation time [18], [19], in a way that the actual time operation can be delayed. In contrast, in the DNN based RA, the generic solver is obtained autonomously through DNN which includes only simple matrix operations, so that the suitable RA can be established with a low computational complexity without iteration [20] for any channel condition.

With the advancement of wireless communication systems, the demands for high data rates and link reliability of the communication systems is increased. The uses of multiple antennas technology known as multiple-input-multiple-output (MIMO) [21] is one of the approaches to solve this issue. In [22], this idea is expanded to a decentralized robust precoding scheme in a MIMO network configuration, which seems like a more challenging setting because of the continuous decision space and the required fine granularity of the precoding, particularly at high SNR.

This paper focuses on applying deep ML methods in end-to-end wireless communication systems. We propose a novel end-to-end wireless communication system architecture utilizing deep ML to overcome significant challenges in deploying this technology in wireless cellular networks. Our approach involves pre-assuming the channel transfer function in deep learning. Initially, we evaluate an end-to-end communication system that jointly optimizes transmitter and receiver elements. We then extend this concept to networks with multiple transmitters and receivers, developing deep ML-based end-to-end systems for MIMO communication. We also introduce a new optimization process tailored for deep ML-based MIMO systems. Our simulations demonstrate that the proposed architecture enables the receiver to learn to recover symbols with minimal errors, thereby enhancing performance.

The rest of the paper is organized as follows: Section II details the system model based on the end-to-end MIMO wireless communication system using deep ML. Section III discusses the optimization process. Section IV presents the simulation results, and Section V concludes the paper.

## II. System Model

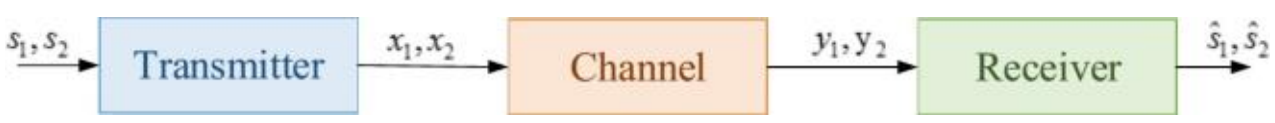

Fig. 1. Communications system model.

The communications system consists of a transmitter, a channel, and a receiver in its simplest form. The simplest form of communication system is shown in Fig. 1. This simple end-to-end single-input single-output (SISO) communication system can be seen as a particular type of *autoencoder* from a DL point of view. Typically, the goal of an autoencoder is to find a low-dimensional representation of its input at some intermediate layer which allows reconstruction at the output with minimal error. The communication rate of autoencoder is $R = k/n$ [bit/channel use], where $k = log_2(M)$. In this paper, $(n, k)$ means that a communications system sends one out of $M = 2^k$ messages (i.e., k bits) by $n$ channel uses. The proposed deep ML-based end-to-end SISO communication system (autoencoder) is shown in Fig. 2.

In this paper, the Rayleigh fading channel in the proposed end-to-end SISO system model is considered. So the transmitter output must be converted to complex samples and $\mathbf{x} \in \mathbb{C}^n$ and $\mathbf{y} \in \mathbb{C}^n$ denote the complex transmit and received signals, respectively. Upon reception of $\mathbf{y}$, the receiver conducts the transformation $g : \mathbb{R}^n \mapsto M$ to create the estimate $\hat{s}$ of the transmitted message $s$. It tries to learn representations $\mathbf{x}$ of the messages $s$ that are powerful regarding the channel impairments mapping $\mathbf{x}$ to $\mathbf{y}$ (i.e., noise, fading, distortion, etc.) so that the transmitted message can be improved with a small probability of error. The decode function is denoted by $g(\mathbf{y})$ and consists of a dense and softmax layer to create the estimate $\hat{s}$. The channel have Rayleigh fading distribution in the proposed deep ML-based end-to-end SISO system. The encoder function is denoted by $f(s)$ and composed of a dense and normalization layer to normalize average power in the system.

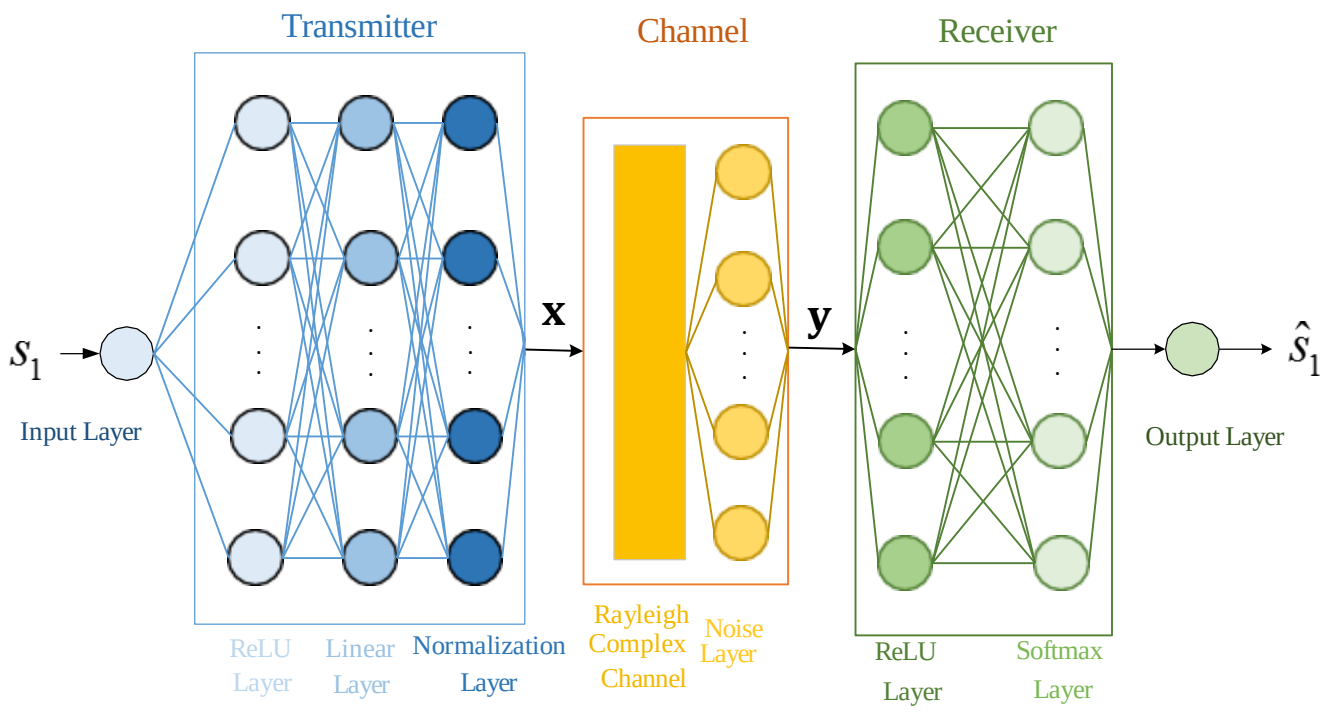

Fig. 2. The deep ML-based SISO communication system over an Rayleigh fading channel.

The training of the end-to-end SISO communication system is shown in Fig. 3. During the training, known transmitted data act as training data, where the transmitted symbols are generated and sent over the channel. During this process, the transmitter, the receiver, and the channel layers can be trained, and the parameters of each layer will be obtained.

Now, we introduce a deep ML-based end-to-end MIMO communication system. Let us consider an uncoded $2 \times 2$ MIMO system (see Fig. 4). The received signal at the $i$-th antenna, $\mathbf{y}_i$, for $i = 1, 2$, can be written as:

$$\begin{aligned} \mathbf{y}_1 &= \mathbf{h}_{1,1}\mathbf{x}_1 + \mathbf{h}_{1,2}\mathbf{x}_2 + n_1 \\ \mathbf{y}_2 &= \mathbf{h}_{2,1}\mathbf{x}_1 + \mathbf{h}_{2,2}\mathbf{x}_2 + n_2 \end{aligned} \tag{1}$$

In matrix form the received signal can be defined as follows:

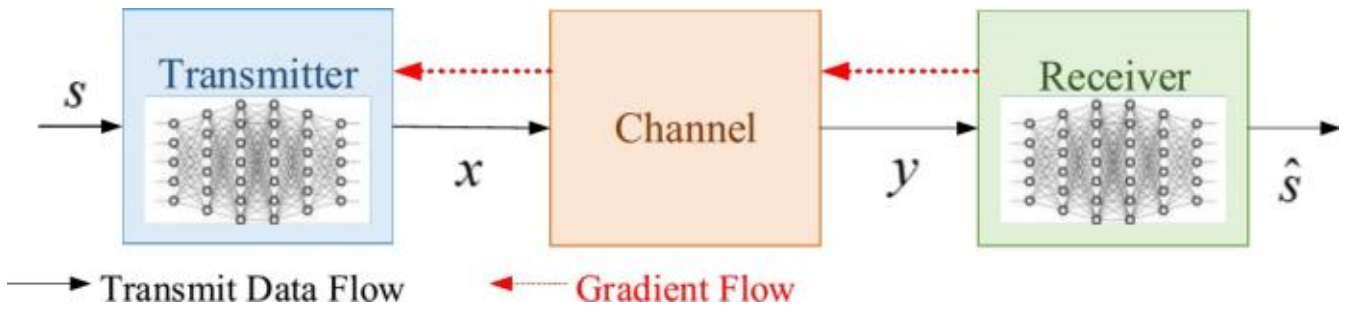

Fig. 3. Training of the end-to-end system.

$$\mathbf{Y} = \mathbf{H}\mathbf{x} + \mathbf{n} \tag{2}$$

where $\mathbf{H}$ is the channel matrix and the channel components are i.i.d. Rayleigh fading channel taps, $\mathbf{x}$ is the transmit signal vector, and $\mathbf{n}$ represents the additive white Gaussian receiver noise vector with zero-mean. The proposed deep ML-based end-to-end MIMO communication system is shown in Fig. 5. Here, the transmitter consists of an feedforward neural network (FNN) with multiple dense layers followed by a normalization layer that ensures that physical constraints on $\mathbf{x}$ are met.

The normalization layer is a simple method to enhance the training speed for different neural network models. This method calculates the normalization statistics directly from the summed inputs to the neurons within a hidden layer so that the normalization does not introduce any new dependencies between training cases. The energy and average power constraints are considered limitations in the normalization layer. As mentioned, the channel is represented by $\mathbf{H}$, where the channel components are i.i.d. Rayleigh fading channel taps. The receiver uses an FNN with multiplex dense layers for signal detection. Its last layer employs a softmax activation whose output $\mathbf{p} \in (0, 1)$ is a probability vector over all possible messages.

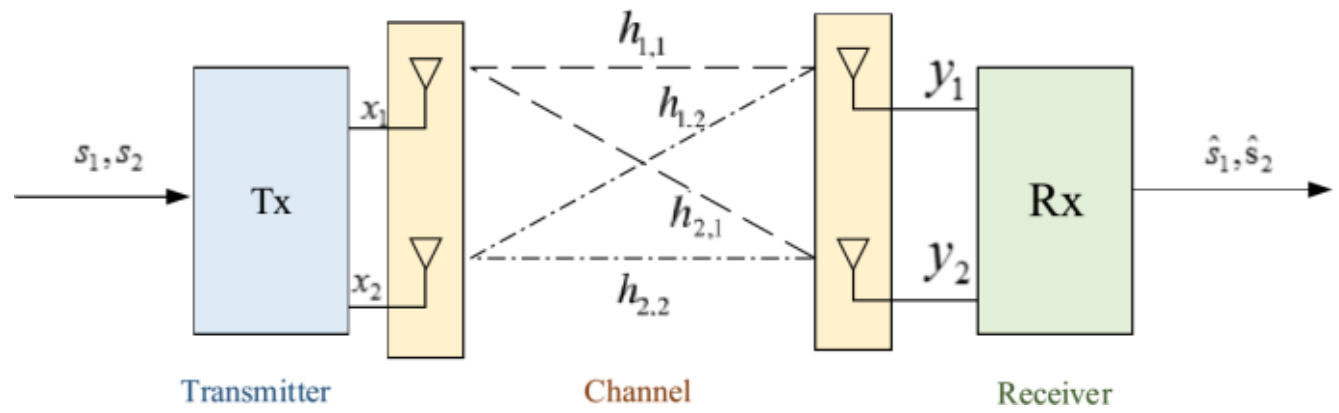

Fig. 4. A simple MIMO communication system.

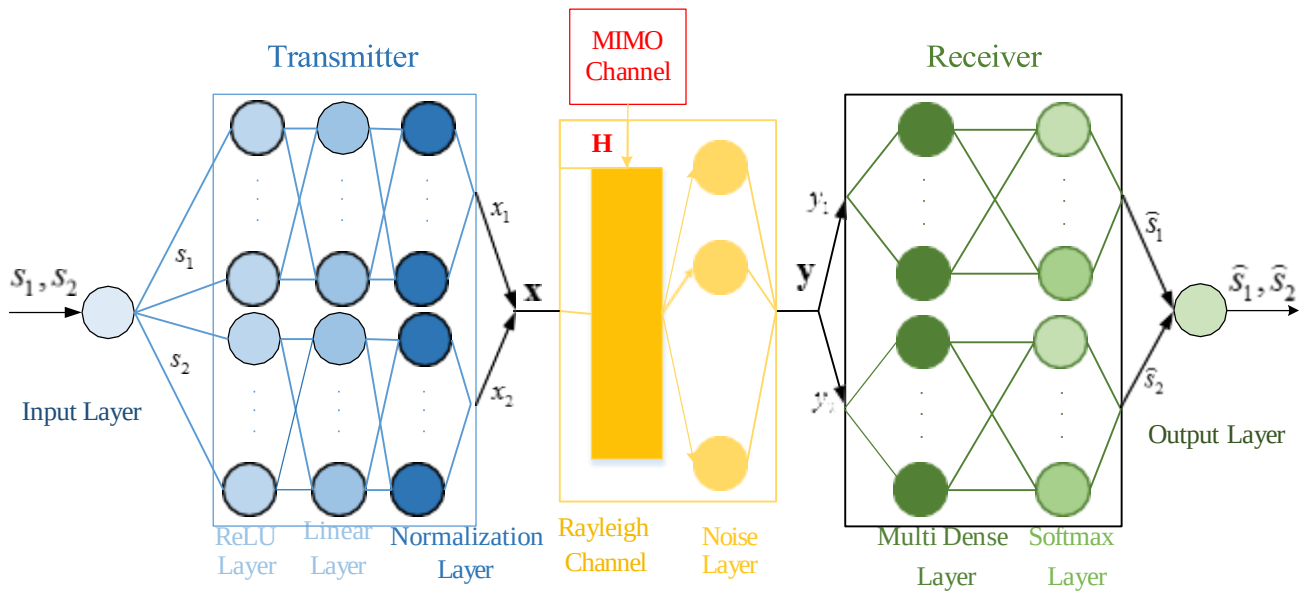

Fig. 5. Proposed end-to-end MIMO wireless communication system using Deep ML.

MIMO communication systems are classified into two groups. The first is open-loop (without any channel state information (CSI) at the transmitter), and the second is closed-loop (with CSI at the transmitter feedback from the receiver). In this paper, we consider the open-loop MIMO communication system. The proposed system encodes $k$ bits (s) to form $m_t$ parallel transmit streams of $n$ time samples. These streams transmit by multi-antenna mixing channel and arrive at a receiver to create $m_r$ receive streams, each of length $n$ time samples and decoded to produce an estimate for $\hat{s}$.

Core to the ability to optimize such an end-to-end system is to accurately model or represent the MIMO channel effects within the network transfer function $\hat{s} = f(s; \theta)$. We use several custom layers to model the domain enumerated below to simulate MIMO propagation. We can express the complete network $f$ as follows for the open loop MIMO encoding case:

$$f(s; \theta) = \sigma_1(\mathbf{W}_l \mathbf{r}_{l-1} + \mathbf{b}_l) \tag{3}$$

where $\mathbf{W}_l \in \mathbb{R}^{N_l \times N_{l-1}}$, $\mathbf{b}_l \in \mathbb{R}^{N_l}$ and $\sigma_1(.)$ is the softmax activation function and $\mathbf{r}_{l-1}$ is the output vector from the previous layer (Dense layer). $\mathbf{r}_{l-1}$ can be defined as follows:

$$\mathbf{r}_{l-1} = \sigma_2(\mathbf{W}_{l-1}\mathbf{r}_{l-2} + \mathbf{b}_{l-1}) \tag{4}$$

where $\mathbf{W}_l \in \mathbb{R}^{N_{l-1}\times N_{l-2}}$, $\mathbf{b}_{l-1} \in \mathbb{R}^{N_{l-1}}$ and $\sigma_2(\cdot)$ is the ReLU activation function and $\mathbf{r}_{l-2} = \mathbf{y}$ is the output vector from the previous layer (Noise layer). $\mathbf{y}$ can be defined as follows:

$$\mathbf{y} = \sigma_3(\mathbf{r}_{l-3}) = \mathbf{r}_{l-3} + \mathbf{n} \tag{5}$$

where $\mathbf{n} \sim \mathcal{N}(0, \beta\mathbf{I}_{N_{l-2}})$ is the additive white gaussian noise and $\mathbf{r}_{l-3}$ is the output vector from the previous layer (Complex matrix multiplication). We defined $\mathbf{r}_{l-3}$ as follows:

$$\mathbf{r}_{l-3} = \mathbf{Hx} \tag{6}$$

where $\mathbf{H}$ is the MIMO channel matrix and $\mathbf{x} = \mathbf{r}_{l-4}$ is the output vector from the previous layer (Normalization layer). $\mathbf{x}$ can be calculated as follows:

$$\mathbf{x} = \sigma_5(\mathbf{r}_{l-5}) = \frac{\sqrt{N_{l-5}}\mathbf{r}_{l-5}}{||\mathbf{r}_{l-5}||_2} \tag{7}$$

where $\mathbf{r}_{l-5}$ is the output vector from the previous layer (Dense layer). $\mathbf{r}_{l-5}$ is defined as:

$$\mathbf{r}_{l-5} = \sigma_6(\mathbf{W}_{l-5}\mathbf{r}_{l-6} + \mathbf{b}_{l-5}) \tag{8}$$

where $\mathbf{W}_{l-5} \in \mathbb{R}^{N_{l-5}\times N_{l-6}}$, $\mathbf{b}_{l-5} \in \mathbb{R}^{N_{l-5}}$ and $\sigma_6(.)$ is the ReLU activation function and $\mathbf{r}_{l-6} = s$ is the system input. We use $\theta = \{\theta_1, ..., \theta_5\}$ to denote the set of all parameters of the network. The $\sigma_1$,$\sigma_3$, $\sigma_5$, and $\sigma_6$ are the Decode, Awgn, Norm and Encode functions, respectively, where

- Decode: Learned Decoder ($y \rightarrow \hat{s}$)
- Norm: Normalize average power
- Channel: Rayleigh fading channel, $\mathbf{H}$
- Multi: Complex matrix multiplication of $\mathbf{x}$ with $\mathbf{H}$
- Awgn: Additive white Gaussian receiver noise with mean zero and variance $\sigma$
- Encode: Learned Encoder ($s \rightarrow x$)

In the Decode, Awgn, Norm, and Encode blocks, we utilize the Dense, Noise, Normalization, and Dense layers, respectively. Using this formulation, forward and backward gradient passes can readily be computed on $f(s; \theta)$. In the backward pass, the Awgn function becomes the identity function (used only for the forward pass). While the normalization module enforces a constant average power, the noise standard deviation $\beta\mathbf{I}_{N_{l-2}}$ may be easily adjusted at training or test time to simulate varying levels of signal-to-noise ratio (SNR).

## III. Optimization Process

In our optimization process, we represent $s$ as a $2^k$ valued integer of codeword indices the system may transmit, each encoding $k$ bits. In the network, we present this as a one-hot input vector of length $2^k$ with a single non-zero value of 1 for the desired codeword and the output as a softmax classification task approximating each codeword's probability.

*1) Deep ML-based SISO:* In the deep ML-based SISO communication system, the loss function of the information symbols is denote by:

$$l_i = -log([\hat{s}_i]_{s_i}). \tag{9}$$

The categorical cross-entropy loss function ($l_T$) may be readily used for optimization using gradient descent to select network parameters. In this case ($l_T$) is given by:

$$\begin{aligned} l_T(s, \hat{s}) &= -\frac{1}{m}\sum_{i=0}^{m-1} l_i \\ &= -\frac{1}{m}\sum_{i=0}^{m-1} (s_i log(\hat{s})) \end{aligned} \tag{10}$$

where $m$ is the number of codewords used during training. Using a form of stochastic gradient descent, the weight updates can be computed using backpropagation. In this case, we iteratively compute a forward path: $\hat{s} = f(s, \theta)$, and a backward path:

$$\nabla L = \frac{\partial l_T(s, f(s, \theta))}{\partial \theta} \tag{11}$$

where network layer weights are given by $\theta$ and a weight update takes the form of $-\eta\nabla L$ and $\eta$ represents a learning rate.

*2) Deep ML-based MIMO:* In this paper, we minimizing a weighted sum of loss functions to find system parameters. In the deep ML-based MIMO communication system, the loss functions of the first and second transmitter-receiver pair are denote by:

$$l_1 = -log([\hat{s}_1]_{s1}) \quad l_2 = -log([\tilde{s}_2]_{s2}), \tag{12}$$

and $\tilde{L}_1(\theta_t)$ and $\tilde{L}_2(\theta_t)$ are the associated losses for mini-batch $t$. In this paper, we utilize cross-entropy loss functions. In such a context, it is less clear how one should train with conflicting goals. One approach consists of minimizing a weighted sum of losses. The weighted sum of losses is denoted by:

$$\tilde{L} = \gamma \tilde{L}_1(\theta_t) + (1-\gamma)\tilde{L}_i(\theta_t) \tag{13}$$

where $\gamma \in [0,1]$ and we givie equal weight to both losses.

## IV. Numerical Results

In this section, we describe results from two preliminary studies that directly support the proposed idea in this research. We evaluate our system model in Python and plot our results in Matlab. As mentioned in our simulation, we consider the channel to be a Rayleigh fading channel tap in the proposed ML-based end-to-end communication system, so the transmitter output must be converted to complex samples.

*1) Deep ML-based SISO:* First, we evaluate the deep ML-based end-to-end SISO communication system model. As mentioned, the simple end-to-end SISO communications system is an autoencoder. In our simulation, the communication rate of the autoencoder is $R = 4/7$ [bit/channel use], where $k = 4$. In the sequel, we consider the notation $(n, k) = (7, 4)$ means that a communications system sends one out of $M = 2^k = 2^4$ bit messages through 7 channel uses. To simulate conventional SISO, we consider the Hamming block code and BPSK modulation to convert bits to symbols in the transmitter. The layouts of the deep ML-based end-to-end SISO communication system are provided in Table I.

TABLE I
Layout of Authoencoder Used in Proposed End-to-End SISO Systems

| Layer | Output dimensions |
|---|---|
| Input | 2M |
| ReLU | 2M |
| Linear | 2n |
| Normalization | 2n |
| Noise | 2M |
| ReLU | 2M |
| Softmax | 2M |

In other words, while most autoencoders eliminate redundancy from input data for compression, this autoencoder often adds redundancy, performing an intermediate representation robust to channel perturbations. The BPSK modulation scheme is employed in the simulation to modulate the input bits for the Python simulations, and Rayleigh fading is utilized as the channel model. The input $s$ is encoded as a one-hot vector, and the output is a probability distribution over all possible messages from which the most likely is picked as output $\hat{s}$. The simulation results of the proposed deep ML-based end-to-end SISO communication system are illustrated in Fig. 6. We show that in the deep ML-based end-to-end SISO system, the receiver in the testing step manages to recover the symbols with few errors. This figure evaluated the proposed deep ML-based end-to-end SISO communication system over the Rayleigh fading channel. We compared the proposed method with the conventional SISO communication system over Rayleigh fading. Also, we showed the coded and uncoded BPSK modulation results over AWGN and Rayleigh fading channels.

*2) Deep ML-based MIMO:* We now evaluate the performance of the deep ML-based $2 \times 2$ end-to-end MIMO communication system that uses spatial multiplexing to increase throughput with a single time slot using both a conventional and an end-to-end MIMO communication system. The channel is assumed to be Rayleigh fading, and equal power is used at each antenna during transmission for both the conventional MIMO and the deep ML-based end-to-end MIMO systems. QPSK modulation encodes each data stream for conventional and deep ML-based end-to-end MIMO communication systems. The layouts of deep ML-based end-to-end MIMO communication systems are provided by replacing $2n$ with $4n$ and $2M$ with $4M$ in Table I. In conventional MIMO, we used the maximum likelihood technique in the receiver. Simulation result of the proposed deep ML-based end-to-end MIMO communication system shown in Fig. 7.

## V. Conclusion

This paper introduced and evaluated a deep machine learning-based end-to-end MIMO communication system optimized using SGD with backpropagation of the loss function gradients from the output to the input layer. By pre-assuming the channel transfer function, we developed autoencoders tailored to SISO and MIMO communication systems operating over Rayleigh fading channels to minimize BER. Our results demonstrate that the proposed system exhibits competitive performance, particularly in high SNR scenarios, effectively learning to recover symbols with minimal errors. This highlights the potential of deep learning frameworks to enhance the efficiency and reliability of modern wireless communication systems, offering a promising alternative to traditional methods.

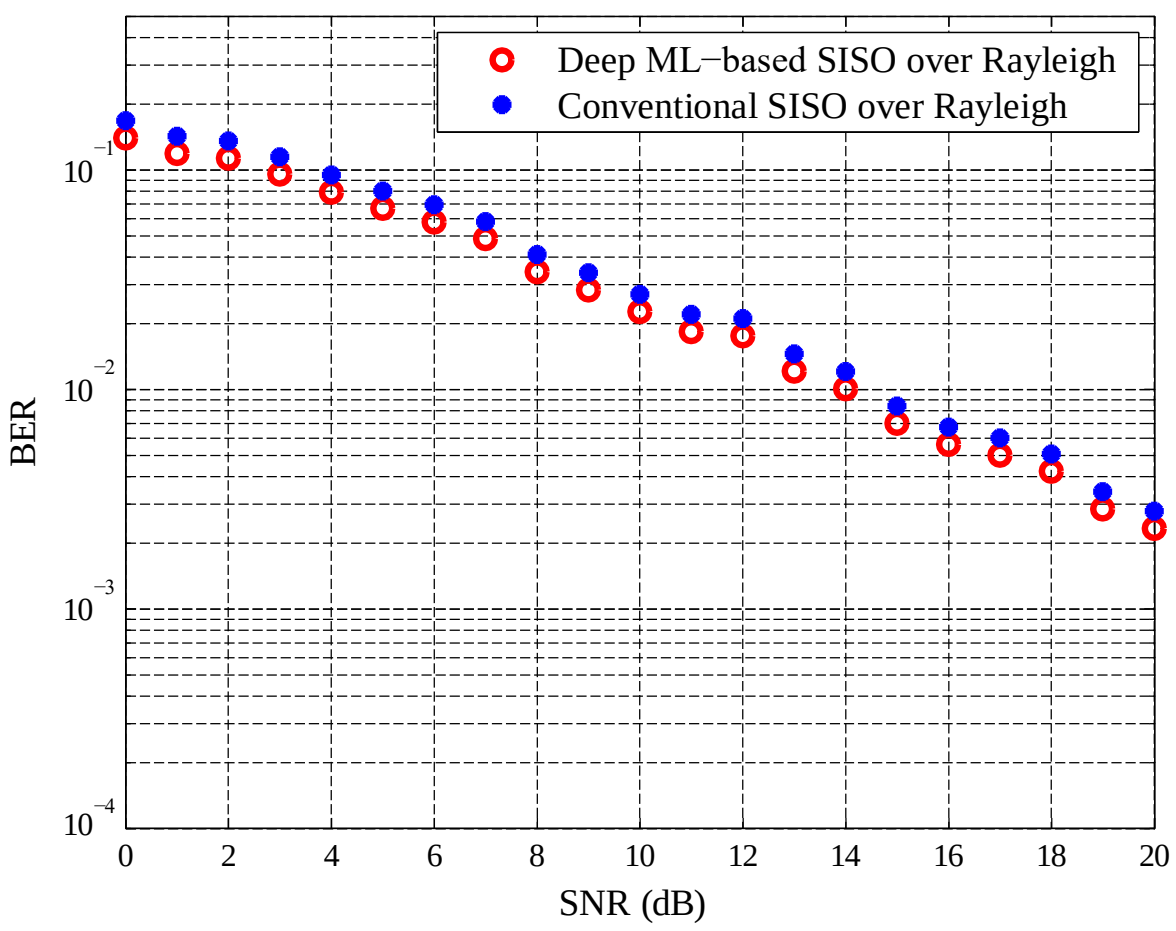

Fig. 6. BER performance of the deep ML-based SISO and several communication schemes.

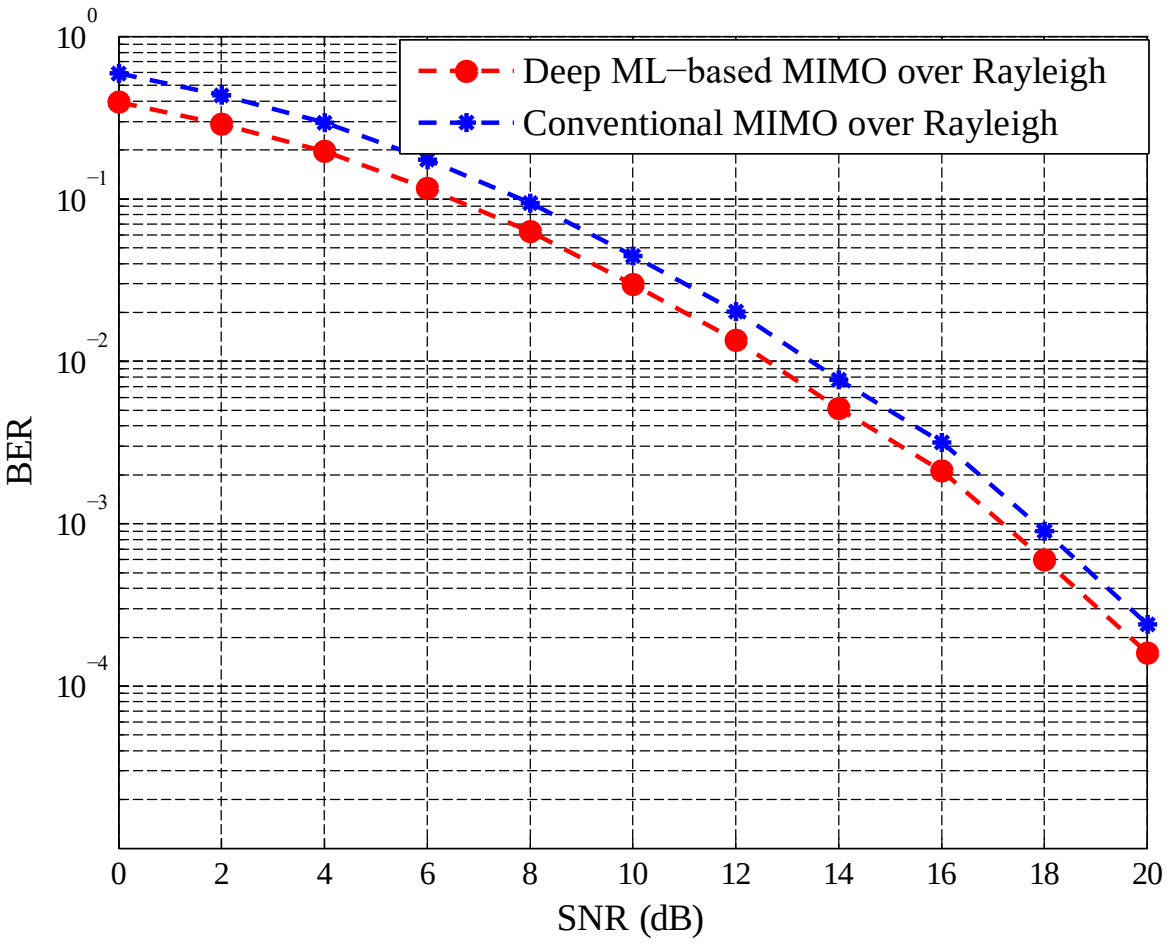

Fig. 7. BER performance of $2 \times 2$ conventional and deep ML-based MIMO communication schemes.